\documentclass[graybox]{svmult}

\usepackage{type1cm}        

\usepackage{amsfonts,amssymb,amsbsy,tabulary,amsmath}
\smartqed  
\usepackage{graphicx}
\usepackage{ulem}
\usepackage{amsmath}
\usepackage{natbib}
\usepackage{makeidx}         
\usepackage{graphicx}        
\usepackage{multicol}        
\usepackage[bottom]{footmisc}

\usepackage{newtxtext}       %
\usepackage[varvw]{newtxmath}       

\usepackage{makecell}
\usepackage{url,multirow,morefloats,floatflt,cancel,tfrupee}
\makeatletter

\AtBeginDocument{\@ifpackageloaded{textcomp}{}{\usepackage{textcomp}}}
\makeatother
\usepackage{colortbl}
\usepackage{xcolor}
\usepackage{pifont}
\usepackage{doi}

\usepackage[nointegrals]{wasysym}
\makeatletter

\def\mcWidth#1{\csname TY@F#1\endcsname+\tabcolsep}

\def\cAlignHack{\rightskip\@flushglue\leftskip\@flushglue\parindent\z@\parfillskip\z@skip}
\def\rAlignHack{\rightskip\z@skip\leftskip\@flushglue \parindent\z@\parfillskip\z@skip}

\@ifundefined{etal}{}{}

\usepackage{ifxetex}
\ifxetex\else\if@twocolumn\@ifpackageloaded{stfloats}{}{\usepackage{dblfloatfix}}\fi\fi

\AtBeginDocument{
\expandafter\ifx\csname eqalign\endcsname\relax
\def\eqalign#1{\null\vcenter{\def\\{\cr}\openup\jot\m@th
  \ialign{\strut$\displaystyle{##}$\hfil&$\displaystyle{{}##}$\hfil
      \crcr#1\crcr}}\,}
\fi
}

\AtBeginDocument{%
  \@ifpackageloaded{endfloat}%
   {\renewcommand\efloat@iwrite[1]{\immediate\expandafter\protected@write\csname efloat@post#1\endcsname{}}}{\newif\ifefloat@tables}%
}%

\def\BreakURLText#1{\@tfor\brk@tempa:=#1\do{\brk@tempa\hskip0pt}}
\let\lt=<
\let\gt=>
\def\processVert{\ifmmode|\else\textbar\fi}

\@ifundefined{subparagraph}{
\def\subparagraph{\@startsection{paragraph}{5}{2\parindent}{0ex plus 0.1ex minus 0.1ex}%
{0ex}{\normalfont\small\itshape}}%
}{}

\newcommand\role[1]{\unskip}
\newcommand\aucollab[1]{\unskip}
  
\@ifundefined{tsGraphicsScaleX}{\gdef\tsGraphicsScaleX{1}}{}
\@ifundefined{tsGraphicsScaleY}{\gdef\tsGraphicsScaleY{.9}}{}
\def\checkGraphicsWidth{\ifdim\Gin@nat@width>\linewidth
	\tsGraphicsScaleX\linewidth\else\Gin@nat@width\fi}

\def\checkGraphicsHeight{\ifdim\Gin@nat@height>.9\textheight
	\tsGraphicsScaleY\textheight\else\Gin@nat@height\fi}

\def\fixFloatSize#1{}
\let\ts@includegraphics\includegraphics

\def\inlinegraphic[#1]#2{{\edef\@tempa{#1}\edef\baseline@shift{\ifx\@tempa\@empty0\else#1\fi}\edef\tempZ{\the\numexpr(\numexpr(\baseline@shift*\f@size/100))}\protect\raisebox{\tempZ pt}{\ts@includegraphics{#2}}}}

\AtBeginDocument{\def\includegraphics{\@ifnextchar[{\ts@includegraphics}{\ts@includegraphics[width=\checkGraphicsWidth,height=\checkGraphicsHeight,keepaspectratio]}}}

\DeclareMathAlphabet{\mathpzc}{OT1}{pzc}{m}{it}

\def\URL#1#2{\@ifundefined{href}{#2}{\href{#1}{#2}}}

\def\UrlOrds{\do\*\do\-\do\~\do\'\do\"\do\-}%
\g@addto@macro{\UrlBreaks}{\UrlOrds}

\edef\fntEncoding{\f@encoding}

\makeatother

\newif\ifmultipleabstract\multipleabstractfalse%
\usepackage[T1]{fontenc}

\makeatletter	

\AtBeginDocument{%
  \@ifpackageloaded{endfloat}%
   {
\renewcommand*\efloat@process[2]{%
  \ef@ifct{#1}{%
    \expandafter\immediate\expandafter\closeout\csname efloat@post#1\endcsname
    \ef@setct{#1}{0}%
    \clearpage                                                         
        
    \efloat@ifflag{#2list}{
      {\normalsize\efloat@listof{#2}}
    }{}%

    \efloat@ifflag{#2head}{%
      \section*{\@nameuse{#2section}}
      \suppressfloats[t]
    }{}

    \markboth                                                          
      {\expandafter\uppercase\expandafter{\csname #2section\endcsname}}
      {\expandafter\uppercase\expandafter{\csname #2section\endcsname}}

    \def\efloat@type{#2}%
    \processdelayedfloat@hook
    \@nameuse{process#2s@hook}%
    \clearpage
    \@input{\jobname.#1}%
  }{}}
   
   }{}%
}%

  \makeatother

\newcommand{\ME}{$\text{M}_{\oplus}$} 
\newcommand{\RE}{$\text{R}_{\oplus}$} 
\newcommand{\ms}{m\,s$^{-1}$}
\newcommand{\cms}{cm\,s$^{-1}$}

\makeindex             

\begin{document}

\title{Origin and characterization of super-Earths and sub-Neptunes}
\author{Léna Parc$^1$, Julia Venturini$^1$, François Bouchy$^1$, Ravit Helled$^2$, Caroline Dorn$^3$,  Adrien Leleu$^1$, Yann Alibert$^{4,5}$, Simon Müller$^2$ and Haiyang Wang$^{3,6,7}$}
\institute{1: Observatoire de Genève, Université de Genève, Chemin Pegasi 51, 1290 Versoix, Switzerland. \\
2: Department of Astrophysics, University of Zurich, Winterthurerstrasse 190, CH-8057 Zurich, Switzerland. \\
3: Institute for Particle Physics and Astrophysics, ETH Zürich, Otto-Stern-Weg 5, 8093 Zürich, Switzerland. \\
4: Center for Space and Habitability (CSH), Universität Bern, Gesellschaftstrasse 6, 3012 Bern, Switzerland. \\
5: Space Research and Planetary Sciences, Physics Institute, University of Bern, Gesellschaftsstrasse 6, 3012 Bern, Switzerland. \\
6: Institute of Geochemistry and Petrology, Clausiusstrasse 25, 8092 Zürich, Switzerland. \\
7: Centre for Star and Planet Formation, Globe Institute, University of Copenhagen, Øster Voldgade 5-7, 1350 Copenhagen, Denmark. \\
}
%
%
\setcounter{chapter}{11}
\titlerunning{Origin and characterization of super-Earths and sub-Neptunes}
\authorrunning{Parc et al.}
\maketitle

\abstract{Super-Earths and sub-Neptunes represent the most common class of exoplanets discovered to date in our galaxy, yet they have no direct analogues in the Solar System. Since 2014, researchers within the NCCR $PlanetS$ have made significant contributions to understanding the origin and nature of these small planets. This chapter provides an overview of the progress made in their detection, characterization, and theoretical interpretation during the 2014–2025 period. The combined data from space-based photometric missions such as Kepler and TESS, together with ground-based radial velocity campaigns using state-of-the-art spectrographs (e.g., HARPS, ESPRESSO, NIRPS), have enabled detailed demographic analyses of these planets. These observational efforts are complemented by theoretical work exploring their internal structures, bulk compositions, formation and evolution, shedding light on the physical processes responsible for the observed diversity. As high-precision observations from facilities like JWST begin to probe the atmospheric composition of individual planets, a more complete picture of super-Earth and sub-Neptune origins is emerging, one that continues to challenge and refine current planet formation theories.}


\section{Introduction}


Over the past few decades, the discovery of exoplanets has revolutionized our understanding of planetary systems. Since the first detection of an exoplanet orbiting a Sun-like star in 1995 \citep{Mayor1995}, nearly 6000 exoplanets have been confirmed using various detection techniques. One of the most striking findings from this era, particularly from NASA’s Kepler mission (2009-2018), is the prevalence of planets with sizes between Earth and Neptune, often referred to as super-Earths and sub-Neptunes. Statistical analyses suggest that over half of all Sun-like stars host at least one of these close-in, sub-Neptune-sized planet \citep[e.g.,][]{Batalha2013,Petigura2013,Marcy2014,Fulton2018,He2019}. This result represents a major paradigm shift in planetary science, as such planets are completely absent from our own Solar System.

The detection and characterization of super-Earths and sub-Neptunes has been made possible by the development of both space- and ground-based observational facilities, with notable contributions from the NCCR $PlanetS$, as detailed in Section \ref{sec:obs}. These advancements have enabled statistical analyses of large planet samples, leading to more robust conclusions. Section \ref{sec:demo} highlights several demographic studies of super-Earths and sub-Neptunes carried out within $PlanetS$.

\begin{figure}[t]
\centering
\includegraphics[width=0.7\linewidth]{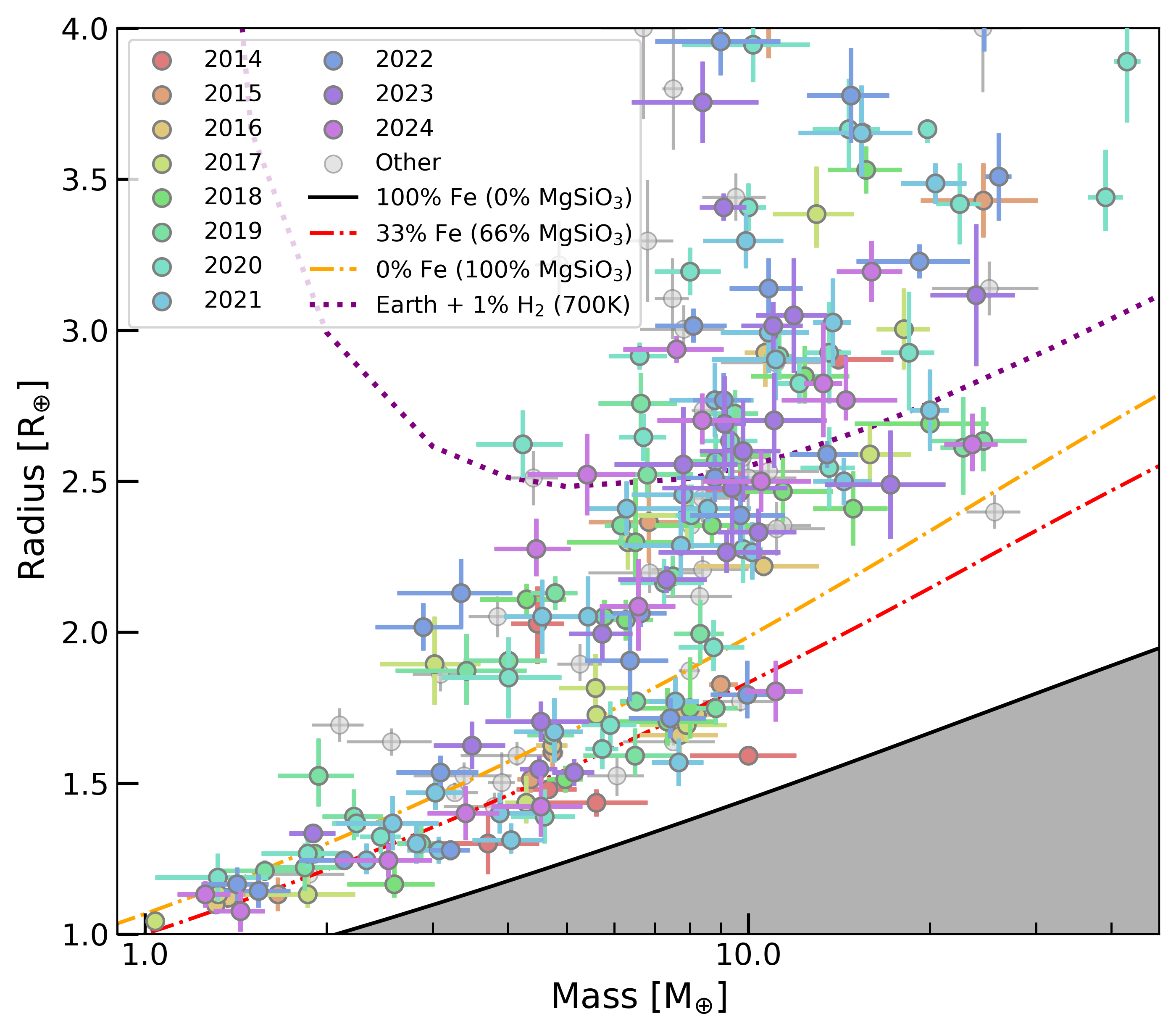}
\caption{Mass-Radius (MR) diagram of small planets from the $PlanetS$ catalog \citep{Otegi2020a,Parc2024} color-coded by their year of discovery using \texttt{mr-plotter} (https://github.com/castro-gzlz/
mr-plotter/). The composition lines of pure-silicates (yellow), Earth-like planets (red), pure-iron (black) and Earth-like+1\% H-He atmosphere (purple) from \citet{Zeng2019} are displayed.
}
\label{fig:MR_year}
\end{figure}

A key demographic breakthrough occurred in 2017, when improved stellar parameters enabled more accurate planet radius estimates. This affected the estimation (and precision) of the planetary radii and allowed to infer that the size distribution of small exoplanets with orbital periods within 100 days is bimodal, with a peak at 1.3 and 2.4 Earth radius \citep{fulton_california-kepler_2017}, and a "valley" or "gap" in between at about 1.8 Earth radius for Sun-like stars (see Fig.\ref{fig:MR_year}). Since then, the planets belonging to the peak at 1.3 Earth radius are commonly referred as "super-Earths" and the planets falling on the second peak, at 2.4 Earth radii, as sub-Neptunes. The "valley" or "gap" in between is now typically called the "radius valley".
Notably, the location of the radius valley varies with orbital period but also with the mass
of the host star \citep{Fulton2018,Berger2020,Ho2023}. 
Several theories have been proposed to explain the origin of the radius valley, which we discuss in detail in Section \ref{sec:evo}. 

Together with understanding their formation and evolution, characterizing the physical nature of these planets has become a central objective. Sub-Neptunes are not expected to have a fully rocky composition \citep{rogers_most_2015,fulton_california-kepler_2017}, and they fall within a region of the mass–radius (M–R) parameter space that is known for its inherent compositional degeneracy. Their bulk densities can be interpreted in multiple ways. For instance, they may consist of solid, rocky cores surrounded by extended hydrogen-helium envelopes, often referred to as “gas dwarfs” \citep{lopez2014understanding,Rogers2023}. Alternatively, they could contain significant amounts of water, either within their interiors or in the form of thick steam or supercritical water atmospheres, commonly termed “water-worlds” \citep{Leger2004,Dorn&Lichtenberg,Aguichine2021exopl,Luque2022}. Section \ref{sec:compo} highlights recent developments in modeling the internal structure and composition of these planets.


This chapter offers a non-exhaustive overview of the advances in our understanding of super-Earths and sub-Neptunes since the launch of $PlanetS$ in 2014, highlighting key scientific contributions made within the NCCR $PlanetS$ framework between 2014 and 2025. 

\section{Observations, detection and characterization of super-Earths and sub-Neptunes}
\label{sec:obs}

Since 2014, the number of discovered super-Earths and sub-Neptunes has drastically increased. Considering transiting exoplanets with both mass and radius reliably measured from the catalog $PlanetS$ (see section \ref{sec:demo}), 
the number of exoplanets with a radius below 4 {\RE} increased from 43 in 2014 to 225 in 2025 (see Fig.~\ref{fig:MR_year}). This rise of new transiting super-Earths and sub-Neptunes was made possible thanks to new space- and ground-based photometric facilities as well as new radial-velocity (RV) facilities. 

Before the start of $PlanetS$ in 2014, most of the known small exoplanets (with $R \lt 4 ${\RE})  were detected by the Kepler space mission \citep{Borucki2010}. Kepler provided an indisputable confirmation of the existence and high occurrence of small-size planets in tight orbits. However, the faintness of the Kepler host stars has been hampering a comprehensive characterization of these systems through ground-based follow-up. Most of them are too far away and too faint for any meaningful follow-up work, and only a small number of them have been adequately characterized. This situation was improved in 2014 with the advent of the NASA K2 mission that uses the repurposed Kepler spacecraft \citep{Howell2014} following the loss of pointing stability that occurred in 2013. K2 scanned new stellar fields along the ecliptic plane until end 2018, changing fields every $\sim$80 days and identifying new transiting planets, orbiting typically brighter stars than the original Kepler discoveries. In 2018, the situation drastically changed with the NASA TESS mission \citep{Ricker2015}. Selected by NASA and lead by MIT, TESS was launch in April 2018 and carried out these 7 last years an almost all-sky photometric survey. TESS successfully identified hundreds of new small-size transiting exoplanets orbiting much brighter stars than the original Kepler discoveries. Launched in 2019, the Swiss-led first ESA small-class mission CHEOPS \citep{Benz2021} (see the chapter 14 ``PlanetS contributions to the CHEOPS mission'') is intensively used to refine, whenever possible, existing radii measurements, and to determine the true orbital period of mono- or duo-transit TESS candidates. New ground-based photometric facilities have also been developed, maintained, or operated with the support of $PlanetS$ these last years and intensively used for the follow-up of TESS candidates. EulerCam is a 4k x 4k CCD detector installed at the Cassegrain focus of the 1.2 m Euler-Swiss telescope at the ESO La Silla site. EulerCam is commonly used for the photometric follow-up of K2 and TESS candidates to refine their ephemeris and radius \citep[e.g.,][]{Lendl2017}. The New-Generation Transit Survey (NGTS; \cite{Wheatley2018}) is a wide-field (96 deg$^2$) photometric survey operating at the ESO Paranal Observatory since 2016. Using twelve independent mounted 20-cm telescopes, NGTS is routinely achieving a precision below 1 mmag. ExTrA is composed of three 60-cm telescopes operating at La Silla Observatory \citep{Bonfils2015}. 
It is dedicated to searching for exoplanets transiting nearby M dwarfs with near-infrared photometry. The instrument relies on a new approach that involves combining photometry with near-infrared  spectroscopic information in order to mitigate the disruptive effect of Earth’s atmosphere, as well as effects introduced by instruments and detectors. SAINT-EX \citep{Gomez2023} is a fully robotic 1-m telescope with an Andor Ikon CCD and filters optimized to obtain high-precision light curves operating at the National Astronomical Observatory of Mexico, in San Pedro Mártir since 2019. 

Radial-velocity (RV) facilities developed, maintained, or operated by $PlanetS$ are intensively used for the follow-up and the mass measurement of transiting candidates (see the chapter 19 about ``Recent development in high-precision high-fidelity spectrographs for exoplanet research and characterization''). CORALIE spectrograph \citep{Queloz2000} operating on the 1.2-m Swiss telescope in La Silla is commonly used for the vetting of TESS candidates in support to different HARPS and ESPRESSO programs. HARPS and HARPS-N spectrographs \citep{Mayor2003, Cosentino2012} have become, through the hundreds of discoveries of Super-Earths and sub-Neptunes over the last 20 years \citep[e.g.,][]{Delisle2018,Udry2019,Unger2021}, the reference for high-precision Doppler velocity measurements. Reaching a RV precision slightly better than 1 {\ms}, HARPS and HARPS-N has significantly contributed in populating the mass-radius diagram of low-mass exoplanets with precise mass measurements of candidates previously detected by the space missions. ESPRESSO \citep{Pepe2021} represents a real breakthrough in the field of exoplanets: its 2-magnitude gain and improved RV precision with respect to HARPS-like instruments is enable a thorough exploration of the rocky planet population. ESPRESSO is a super-stable optical high-resolution spectrograph installed at the combined coudé focus of the ESO Very Large Telescope (VLT) at Paranal. With an internal RV precision close to 20 {\cms}, ESPRESSO can detect a super-Earth in the habitable zone of a Sun-like star. ESPRESSO started its science operation in September 2018, when the first TOIs were released \citep[e.g.,][]{Damasso2020}. More recently, the near-infrared spectrograph NIRPS \citep{Bouchy2025} joined HARPS at the ESO 3.6m telescope.  


Since September 2018, $PlanetS$ is strongly involved in the follow-up of TESS candidates targeting different science objectives like : 1) the transition between super-Earth and sub-Neptune populations to improve the understanding of the radius valley, 2) sub-Neptunes with strong stellar irradiation within the so-called Neptune desert, 3) small exoplanets at long orbital periods identified as duo-transit candidates by TESS, 4) small exoplanets transiting M dwarfs, 5) obliquity of low-mass exoplanets. 

Several multi-planetary systems with sub-Neptunes were discovered and their mass characterized:  
the three sub-Neptune transiting TOI-125 including a puffy one, two of them close to the mean-motion resonance (MMR) and two of them with significant eccentricity \citep{Nielsen2020}; the two sub-Neptunes around TOI-1062 close to the mean motion resonance \citep{Otegi2021}; the pair of sub-Neptunes transiting TOI-1064 including a dense one \citep{Wilson2022}; the low-density hot sub-Neptune in the radius valley TOI-755b \citep{Osborn2021}; the highly irradiated dense sub-Neptune TOI-824b on the lower edge of the Neptune desert \citep{Burt2020}.  


Thanks to the overlap of TESS sectors and the photometric follow-up of duo-transit candidates with CHEOPS, warm sub-Neptunes were identified: the low-density sub-Neptune $\nu^2$ Lupi d (2.6 {\RE} and 8.8 {\ME}) with a period of 107.6 days \citep{Delrez2021}; the two warm sub-Neptunes with contrasting densities transiting TOI-815 \citep{Psaridi2024}; the three warm sub-Neptunes close to mean-motion resonances 2:1 and 5:3 around the young star TOI-2076  \citep{Osborn2022}; the two warm Neptunes transiting TOI-1471 \citep{Osborn2023}.



Several small-size exoplanets were also detected around M dwarfs: a super-Earth and a sub-Neptune close to the MMR around the M3 dwarf TOI-1266 \citep{Demory2020}; the sub-Neptune transiting the M2 dwarf TOI-269 with an unusual eccentricity \citep{Cointepas2021}; the three low-density sub-Neptunes just above to the radius valley transiting the M star TOI-663 \citep{Cointepas2024}.




For some specific system, the Rossiter-McLaughlin effect was measured to determine the system spin-orbit angle. HD3167 hosts an ultra-short period super-Earth and a warm sub-Neptune on perpendicular orbits \citep{Bourrier2021}. The eccentric orbit of the sub-Neptune GJ 436b is nearly perpendicular to the stellar equator \citep{Bourrier2018d}. 



Several super-Earth exoplanets located close to the radius valley were discovered: the hot, rocky and warm, puffy super-Earths orbiting TOI-402 \citep{Dumusque2019}; the super-Earths TOI-260b, TOI-286b, TOI-286c and TOI-134b \citep{Hobson2024}. The ultra-short period super-earth 55 Cnc-e was intensively characterized including the refinement of its mass and radius \citep{Bourrier2018b}, the measurement with CHEOPS of its phase curve \citep{Morris2021} and occultation \citep{Demory2023}.








\section{Demographic studies of super-Earths and sub-Neptunes}
\label{sec:demo}


{\it The $PlanetS$ catalog of reliable and well-characterized transiting exoplanets}\\

The $PlanetS$ catalog\footnote{\url{https://dace.unige.ch/exoplanets/}} is a compilation of transiting exoplanets with reliable parameters, initially introduced by \citet{Otegi2020a} with an upper mass limit of 120~$M_\oplus$. Since its first release, the catalog has been expanded to include planets across the entire mass range. The latest update incorporates newly discovered exoplanets, revised parameters for previously known systems, and additional planetary and stellar properties \citep{Parc2024}. 

The $PlanetS$ catalog aims to provide a comprehensive and up-to-date dataset of transiting exoplanets with precise and reliable mass and radius measurements. As in its original version, it relies on the NASA Exoplanet Archive \citep{NEA2025}, which offers one of the most complete and current databases of planetary parameters and references. The selection criteria have remained consistent: only planets with relative uncertainties smaller than 25\% in mass ($\sigma_M/M < 25\%$) and 8\% in radius ($\sigma_R/R <8\%$) are included. Each entry undergoes a rigorous verification process to ensure the robustness and reliability of the photometric and spectroscopic analyses. When multiple sources are available, the most precise and recent reference is chosen, provided that uncertainties are not underestimated. In the $PlanetS$ catalog, most of the planets have their masses derived from radial velocities measurements. For planets with a mass derived by TTVs, we only consider planets whose mass estimations have been shown to be robust against mass and eccentricity degeneracy \citep{Lithwick2012,HaddenLithwick2017,Leleu2023}. In addition, to provide the community a complete catalog in terms of parameters, a cross-match between catalogs was performed to obtain stellar parameters from Gaia DR3 \citep{gaia2016,gaia2023}, as well as orbital obliquities from TEPCat \citep{Southworth2011}. They also included classical calculations in the database – planet bulk density, insolation flux, equilibrium temperature, expected radial velocity semi-amplitude, and transmission/emission spectroscopic metric \citep[TSM/ESM,][]{Kempton2018} – in order to have these values for all planets in a more homogeneous way. As a result, on the 3rd of April 2025, among the 5867 known exoplanets, the $PlanetS$ catalog contains 825 transiting exoplanets with more than 60 parameters. Among these, 225 are small planets with radii below 4~$R_\oplus$. The next update will include information about binarity of host stars (Messamah et al. in prep). 

The $PlanetS$ catalog is built upon two fundamental parameters that are crucial for exoplanet characterization: planetary mass (M) and radius (R). Knowledge of the planetary mass and radius can be used to infer the planetary density, and therefore, its possible composition. These two fundamental properties are measured using different detection methods and in some cases only one of the two is available. It is therefore valuable to understand the relation between planetary mass and radius, also known as the mass-radius (M-R) relation. Given the diversity of planets, it is also important to understand how the M-R relation depends on the planetary type and identify the transitions between them.  \\

{\it Mass-Radius relationships}\\

Numerous studies have attempted to characterize the mass-radius relationship (M-R) of exoplanets, revealing multiple transitions corresponding to different planet types and internal compositions. \citet{Weiss2013} introduced an M-R-incident flux relation with a break at 150~$M_\oplus$, identifying scaling laws of $R \propto M^{0.53}F^{-0.03}$ for planets below this threshold and $R \propto M^{-0.039}F^{0.094}$ above. \citet{hatzes_definition_2015} further examined the mass-density relation, highlighting transitions that span the giant planet to stellar regime.

\citet{bashi_two_2017} fitted the M-R relation empirically, without imposing priors, and identified a transition at $\sim$124~$M_\oplus$ and 12.1~$R_\oplus$. They found $R \propto M^{0.55}$ for small planets and $R \propto M^{0.01}$ for larger ones. Similarly, \citet{chen_probabilistic_2016} used a probabilistic MCMC approach to identify four mass regimes—Terran, Neptunian, Jovian, and stellar—with transitions at 2.04, 132, and 2.66 $\times 10^{4} M_\oplus$. They inferred distinct relations: $R \propto M^{0.28}$ (Terran), $M^{0.59}$ (Neptunian), $M^{-0.04}$ (Jovian), and $M^{0.88}$ (stellar), with results derived solely from data rather than physical models. At the time, few $\sim 1,M_\oplus$ planets were known, so the Terran–Neptunian transition relied on limited data.

\citet{Otegi2020a} focused on planets up to 120~$M_\oplus$ with the first version of the $PlanetS$ catalog and classified them as rocky or volatile-rich using a pure-water composition line. They found $R = 1.03 M^{0.29}$ for rocky planets and $R = 0.70 M^{0.63}$ for volatile-rich ones. 
\citet{edmondson_breaking_2023} similarly divided planets by composition and found a transition from icy planets to gas giants at 115~$M_\oplus$. They reported $R \propto M^{0.34}$ (rocky), $M^{0.55}$ (icy), and for gas giants, a flat mass dependence with $R \propto M^{0.00}T^{0.35}$, introducing an explicit temperature dependence.


\citet{Muller2024} used the updated $PlanetS$ catalog to perform a solely statistical approach to determine the breakpoints in the relations used to define the different planetary regimes and determine the distinct dependencies. They found this classification: small planets below 4.4~$M_\oplus$ ($R \propto M^{0.27}$), intermediate-mass planets between 4.4 and 127~$M_\oplus$ ($R \propto M^{0.67}$), and giant planets above 127~$M_\oplus$ ($R \propto M^{-0.06}$). They also found a transition between "small" and "intermediate-size" planets in the radius-density relation occurring at a radius of $\sim$1.6~$R_\oplus$.

Building on this, \citet{Parc2024} updated the M-R relations using the same $PlanetS$ catalog but introduced a physically motivated classification: rocky planets below 10~$M_\oplus$, and a threshold of 20\% water content at 650~K to separate rocky from volatile-rich planets, following \citet{luo_interior_2024}. Using automated breakpoint detection, they identified the giant planet transition at $138^{+21}_{-42}$~$M_\oplus$. The best-fit M-R relations were $R \propto M^{0.28}$ (rocky), $M^{0.67}$ (intermediate), and $M^{0.01}$ (giant), consistent with empirical observations and theoretical expectations.

\begin{figure}[t]
\centering
\includegraphics[width=0.9\linewidth]{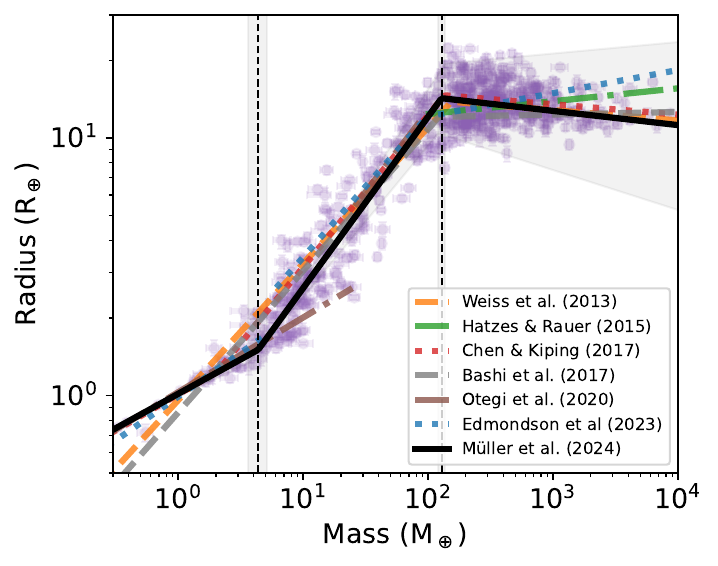}
\caption{\label{fig:MR} Comparison of M–R relations from different studies with exo-planetary data from the PlanetS catalog (Figure from \citet{Muller2024}). 
}
\end{figure}

Figure~\ref{fig:MR} compares the M–R relations from various studies with data from the PlanetS catalog. For rocky and super-Earth exoplanets, all relationships point toward a radius scaling with cubic root of the mass. Mass and radius are not sufficient to determine the transition between rocky and volatile-rich populations. The dispersion of the sub-Neptune population around the M-R relationship is significantly larger than the one in the rocky population. For (sub)-Neptunes, the radius scales with mass to the power 2/3. The transition between volatile-rich population and giant exoplanets arises between the mass of Saturn and half the mass of Jupiter. \\

{\it Multi-planetary systems close to the mean-motion resonance} \\

In multi-planetary system close to the mean-motion resonance (MMR), planetary orbits are influenced and affected by the presence of other planets. 
This modifies in turn the time at which these planets transit, leading to an observable called transit timing variations (TTV). For observation baseline of a few years, TTVs are most pronounced in systems where pairs of planets lie near mean-motion resonances (MMRs), typically when $P_\mathrm{out}/P_\mathrm{in} \approx (k+q)/k$, where $k$ and $q$ are integers, with $q$ typically small ($\leq 2$). Although such near-resonant configurations are relatively uncommon in the overall planet population \citep{Fabrycky2014}, they are disproportionately represented among systems of sub-Neptunes with well-constrained densities \citep{Leleu2024b}, thanks to Kepler's long temporal baseline, as well as dedicated follow-up, notably of K2 and TESS targets. Notably, TTVs enabled the measurement of the masses of the Trappist-1 system with precision better than 5\% \citep{Agol2021}, and sub-10\% precision is obtained more and more for sub-Neptunes with periods $\gtrsim 10$days \citep{Leleu2023,Leleu2024a}.

Planets in resonance are of particular interest, since this orbital configuration is a byproduct of planet migration \citep[e.g.][]{GoTre1979,Weidenschilling1985,Masset2006,TePa2007}. In particular, the occurrence of planets in resonance has been a major tension when comparing the observed population with the outcome of planet formation models in the past decade. However, it has been shown that large TTVs induce a bias against the detection of small planets near resonance \citep{GarciaLopez2011,Kane2019}. Indeed, the algorithms used to detect small planets in transit surveys typically rely on the planet having a periodic orbit, which is by definition not the case when a planet exhibits TTVs. \cite{Carter2013} proposed the detection algorithm QATS that allowed for TTVs, but relied on detecting individual transits, which cannot be reliably done when the signal-to-noise ratio (S/N) of each transit is below 4, which is the case of more than half of the candidates found by the Kepler mission.

In recent years, it was shown that machine learning could be used to recover small planets with large TTVs with individual transits of S/N $\lesssim$ 1 \citep{Leleu2021b,Leleu2022}. These planets were either missed or wrongly flagged as false positives by the pipeline of the Kepler mission (see the chapter 21 ``Machine Learning as a Transformative Tool for (Exo-) Planetary Science'' for more details on these methods). These methods can also be used to better characterize planets whose S/N of individual transit is intermediate ( 2 $\lesssim$ S/N $\lesssim$ 4). Combined with a photo-dynamical analysis \citep{Ragozzine2010}, it enabled to demonstrate that planets in this S/N range had previously under-estimated densities \citep{Leleu2023}. However, the newly determined densities, shown to be robust to the well-known mass-eccentricity degeneracy \citep{Boué2012,LithwickWu2012}, still had lower densities on average than the bulk of sub-Neptunes. 

These results, combined with the gradual increase of well-characterized sub-Neptune, allowed to tackle the tension of the planets whose mass was measured by RVs and TTVs that was discussed since the Kepler mission \citep{Wu2013, Weiss2014, Mills2017, HaddenLithwick2017, Cubillos2017, Millholland2019, Leleu2023, Adibekyan2024}. Indeed, \citet{Leleu2024b} showed that sub-Neptunes at a given distance to resonance had similar densities, regardless of whether they were characterized by RV or TTVs, and that the apparent discrepancies was due to the fact that resonant planets, mainly characterized by TTVs, were puffier than non-resonant planets. They also showed that planets in resonant systems with high multiplicity tended to have similar densities and be coplanar, while planets far from resonances tended to have a larger variety of density, on average denser, and larger mutual inclination. These results are in good agreement with the breaking-the-chain model, in which most of the systems with close-in sub-Neptunes formed in long resonant chains, but most of these chains became unstable after the disc dispersal. The systems that retain their resonant chain remain coplanar, their planets retaining their primordial hydrogen-helium envelopes, while systems that underwent a post-disk instabilities have misaligned orbits with planets that lost part of their atmosphere through collisions \citep{Bean2021,Leleu2024b}. \\

\newpage
{\it Transition between super-Earths and sub-Neptunes across spectral type} \\

Recent studies took advantage of the growing number of well-characterized small planets to separate them according to their host star's spectral type. 
A well-known feature in the radius distribution of small exoplanets is the “radius valley”, first identified by \citet{fulton_california-kepler_2017}. This gap, located around 1.5-2~$R_\oplus$, separates the two populations of super-Earths and sub-Neptunes. In addition to change with the orbital period of the planets, the precise position of the valley varies with the mass of the host star, as shown in subsequent studies \citep[e.g.,][]{Fulton2018,Berger2020,Ho2023}. For example, the latter finds the radius gap to evolve like $R_{\text{valley}} \propto M_\star^{0.231^{+0.053}_{-0.064}}$, meaning it goes to lower radius for lower mass star. Moreover, around M-dwarfs, the existence of this radius valley remains unclear \citep{Cloutier2020,Luque2022,Ho2024,Parc2024}. Instead, \citet{Luque2022} proposed a compositional gap separating rocky from water-rich planets around M-dwarfs. But recent studies found with an increased sample no robust evidence for a separate population of water-worlds around M-dwarfs \citep{Parviainen2023,Parc2024,Dainese2025}. The theoretical aspects of the radius valley are discussed in Sect. 12.5 of this chapter.

\begin{figure}
\centering
\includegraphics[width=1.\linewidth]{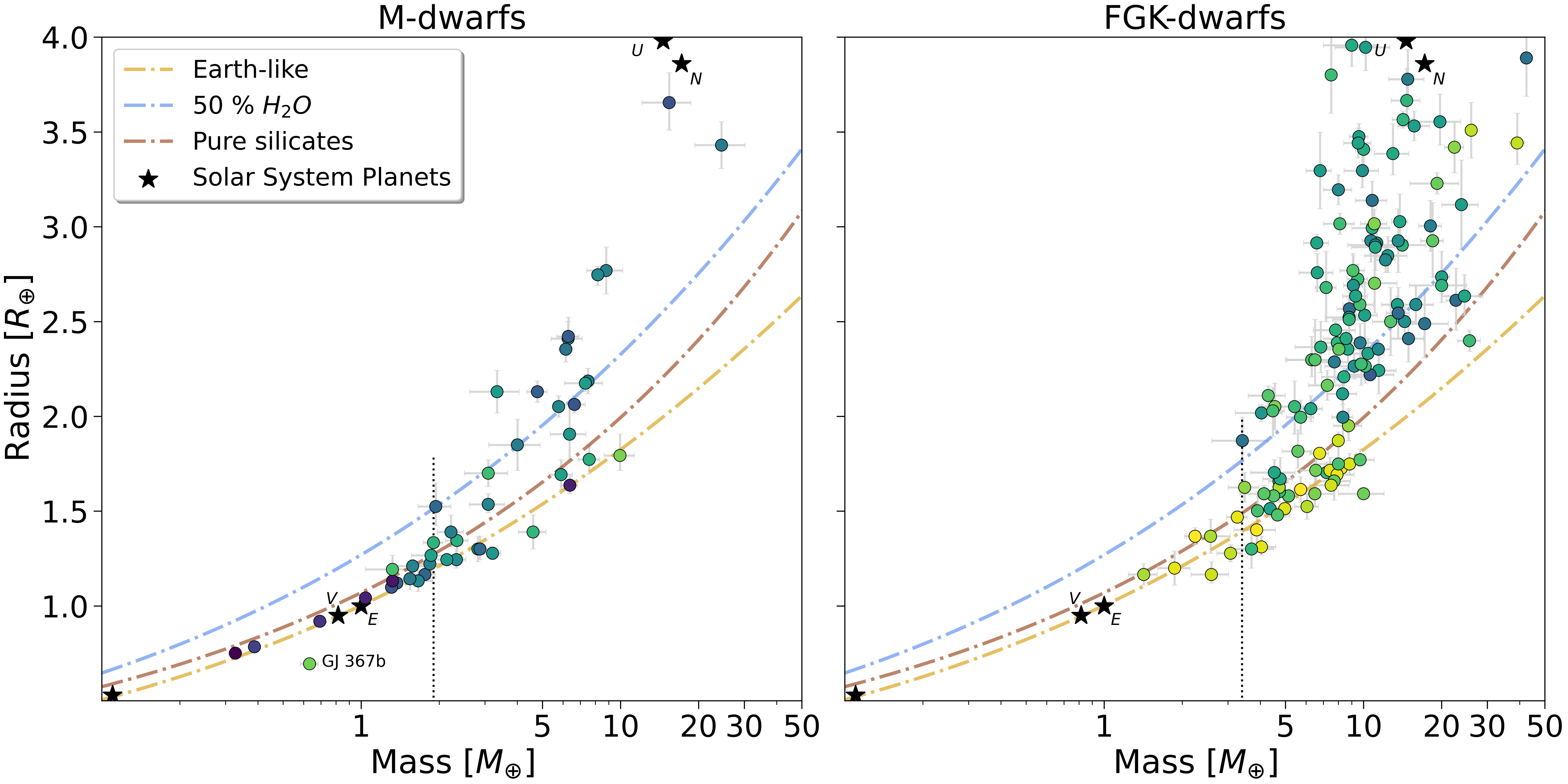}
\caption{\label{fig:MR_Parc24} M-R diagram of small planets around M-dwarfs and FGK-dwarfs from the PlanetS catalog. The planets are color-coded by their equilibrium temperature calculate within the catalog with a bond albedo of 0 and a total heat redistribution. The composition lines of pure-silicates (brown) from \citet{Zeng2016}, Earth-like planets (yellow), and 50\% water (blue) from \citet{Zeng2019} are displayed. The vertical dotted lines correspond to the minimum mass of the sub-Neptunes across spectral types (at 1.9 and 3.4~M$_\oplus$ for M- and FGK-dwarfs) (Figure adapted from \citet{Parc2024}). 
}
\end{figure}

With the $PlanetS$ catalog, \citet{Parc2024} had a closer look to the transition between super-Earths and sub-Neptunes across spectral type. They obtained 46, 61, and 72 planets orbiting M-, K- and FG-dwarfs, respectively (see Fig.~\ref{fig:MR_Parc24}). Their investigation into the super-Earth to sub-Neptune transition reveals that this transition occurs at varying masses and radii contingent upon the host star's spectral type. Notably, while the maximum mass for super-Earths remains around 10$~M_\oplus$ across different stellar types, the minimum mass for sub-Neptunes escalates with the mass of the host star, being approximately 1.9$~M_\oplus$ for M-dwarfs, 3.4$~M_\oplus$ for K-dwarfs, and 4.3$~M_\oplus$ for FG-dwarfs. This trend suggests that planetary migration plays a significant role in shaping the observed distribution, contributing to the less pronounced radius valley for planets around M-dwarfs compared to those orbiting FGK-dwarfs. Additionally, the study finds that smaller sub-Neptunes (1.8$~R_\oplus < R_p < 2.8~R_\oplus$) around M-dwarfs exhibit lower densities than their counterparts around FGK-dwarfs, indicating potential differences in their formation and evolutionary histories. \\

{\it Similarity of Multi-Planetary Systems} \\

The remarkable diversity observed among exoplanets extends to the architectural configurations of multi-planetary systems. The structure of any planetary system is shaped by a complex interplay of physical processes that have governed its evolution, yet the underlying mechanisms responsible for the observed variety in system architectures remain an open question. A particularly striking trend within this diversity is the so-called "peas-in-a-pod" pattern, wherein multi-transiting systems tend to host planets of similar sizes with regularly spaced orbits \citep{Weiss2018}. Understanding these systematic trends is crucial, as they offer valuable insights into the fundamental physical processes driving planetary system formation and place important constraints on theoretical models \citep{Mishra2021}. 
\citet{Otegi2022} utilized the $PlanetS$ catalog to conduct a comprehensive analysis of mass and radius similarities in multi-planetary systems. They also investigated the validity of the "peas-in-a-pod" pattern in terms of planetary density, an aspect that has not been extensively explored in previous studies. They next examined how variations in assumptions and methodological setups commonly employed in the literature influence the observed trends in the "peas-in-a-pod" pattern, particularly with respect to period ratios.
The analysis revealed that while planets within the same system often exhibit similar radii, their masses can vary significantly, and vice versa. However, it was also found  that planetary masses tend to be comparable within a system up to $\sim$ 100 M$_{\oplus}$ and radii of 10 R$_{\oplus}$. Overall, planetary radii within a given system exhibit greater similarity than masses. These findings hint that despite the similar sizes, the compositions and interiors might span a range of possibilities. Moreover, given the inherent diversity among planets within a system, increasing the number of detected multi-planetary systems is essential for advancing our understanding of exoplanetary demographics. \\


\section{Internal structure and composition}
\label{sec:compo}

A key objective of exoplanetary science is to determine the internal structures and bulk compositions of the observed exoplanets. The inferred interior can be used to improve our understanding of planet formation and evolution. This is because the composition and structure are determined by the planetary formation process and its long-term evolution. This topic has been a major focus of research within $PlanetS$.

Generally, the internal structure and composition is derived from transit observations, combined with radial velocity observations. Combining these two types of observations, one can infer the mean density of planets, as well as the composition of the central star. This latter can be related, in term of refractory elements, to the bulk composition of planets \citep{thiabaud2014stellar, dorn2015can, unterborn2017inward,Wang2019Enhanced}, either in a one-to-one relationship, or in a more complex relationship \citep{Adibekyan2021, schulze2021probability,brinkman2024revisiting} (see below).
In principle, there are many bulk compositions and internal structures that can fit the data due to the degenerate nature of the problem \citep{rogers2010framework}. Therefore, our goal is to identify the most probable solutions. In this process, two effects lead to a degeneracy of the solution. First, even for infinitely precise densities, there is an intrinsic degeneracy in the problem (a medium density can be reproduced by either a planet made of pure water, or a planet made of a refractory core surrounded by a lower density gas envelop). This issue is particularly pronounced for super-Earths and sub-Neptunes, which occupy a region of the mass–radius diagram where such degeneracies are especially severe. Observational data alone cannot fully resolve this ambiguity, and only theoretical arguments can reduce it: because we do not imagine a formation path that would lead to the resulting composition, e.g., a pure water planet is implausible. The second effect comes from the observations themselves: even for very precise transit and radial velocity measurements, the planetary radius and mass are measured relative to the stellar values, which themselves can be known to some precision and accuracy only (e.g. using stellar evolution models or astero-seismology in the best cases). Note that, in the case of multiplanetary systems, since by definition planets orbit the same star, the \textit{ratio} of the masses (and radii) of two planets could be determined with very high precision, independently of the precise value of the stellar parameters \citep{Leleu_etal_2021,egger_etal_2024}. More precision regarding the derivation of planetary internal structure using CHEOPS data as an example are presented in chapter 14 of this book. \\

To model the planetary interior, the four structure equations are solved. These include the equations of mass conservation, hydrostatic equilibrium, thermal transport, and energy conservation. We note that the planet is assumed to be in hydrostatic equilibrium and spherically symmetric.  Knowledge of the equations of states (EoSs) of the assumed materials is required to determine and link the thermodynamic properties. \\

{\it Interior-atmosphere exchange}\\

\begin{figure}
\centering
\includegraphics[width=.4\linewidth]{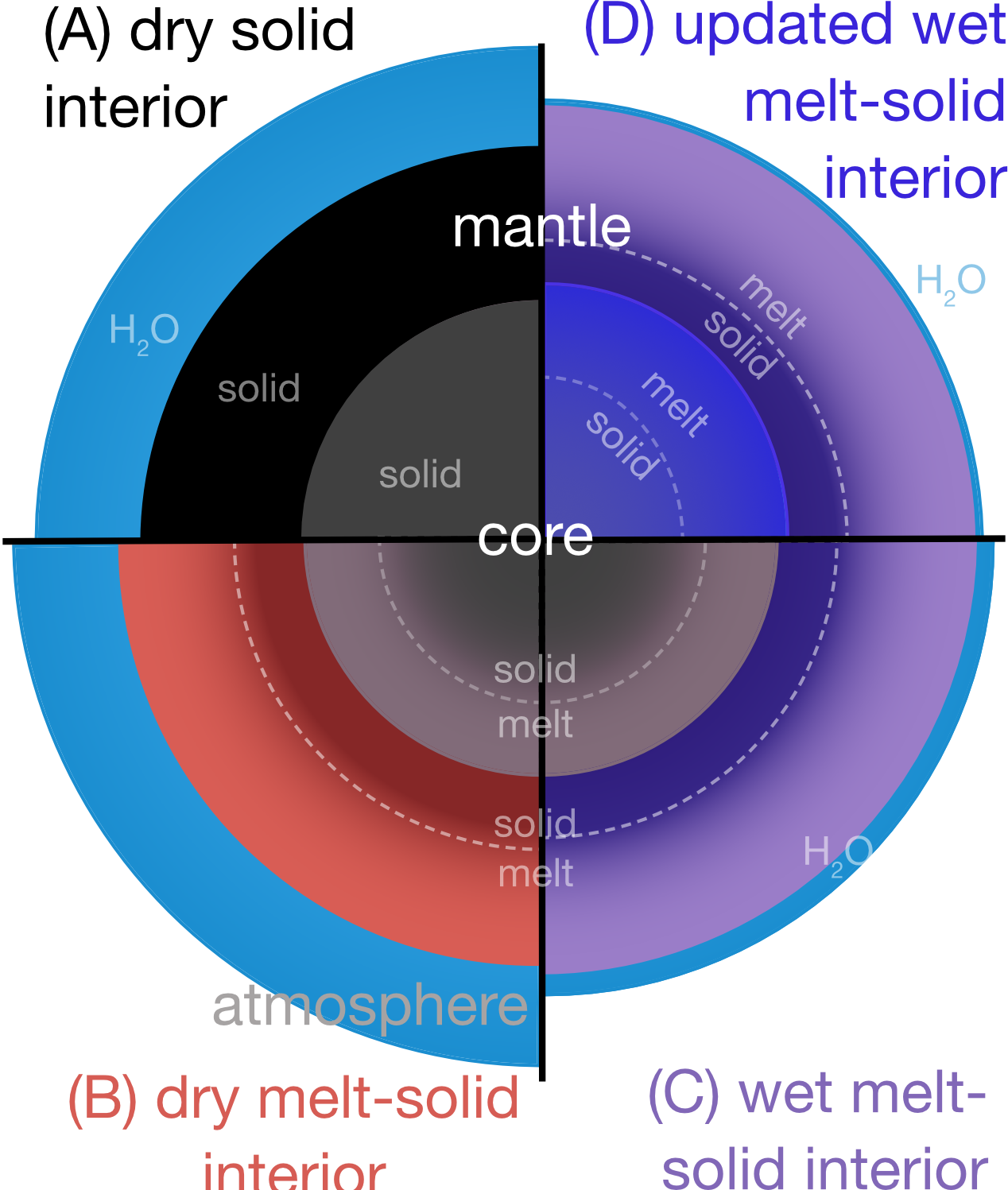}
\caption{\label{fig:sketchplanets} Effect of water dissolution on planetary radius (adapted from \citet{luo_interior_2024}). All interiors share identical bulk composition but differ in model assumptions. Classical models separated volatiles and refractories, and ignored melts (A). Alternative models that include melts but neglected volatile dissolution  overestimate planetary radii (B). The water dissolution in the mantle (C) or mantle and core (D) reduces them. Scenario (D) is considered most realistic for hot super-Earths with magma oceans.
}
\end{figure}

In the past, interior models were assumed to be chemically inactive with separate layers on top of each other. Geochemistry however demonstrate that for the conditions in super-Earths and mini-Neptunes, exchange processes between the gas and the refractory components of planets take place. Most prominent findings in the recent years stem from global chemical equilibrium models that put constraints on plausible bulk compositions of planets \citep{chachan2018role,schlichting2021chemical}.
Chemical interactions between hydrogen-dominated envelopes and magma oceans can generate up to several weight percent of water \citep{Rogers2024} for planets built from dry material. This mechanism of water production may be especially important for rocky super-Earths, which likely formed with hydrogen envelopes that were later lost \citep{rogers2025most}, though it could also have contributed to Earth’s early evolution \citep{young2023earth}. However, planets that acquire water primordially do not necessarily end up with water-rich envelopes. A significant fraction of this water can be sequestered into the deep planetary interior, leaving envelopes depleted in water \citep{Dorn2021, luo_interior_2024}. This has significant impact on calculated radii (see Figure \ref{fig:sketchplanets}). Consequently, even planets with abundant initial water inventories typically retain only a few percent of their mass as water, challenging the common assumption that ice accretion naturally produces water-rich atmospheres \citep{werlen2025sub}.

{\it Impact of measured parameters of exoplanets on the inferred internal structure}\\

Recent work has moved from qualitative statements about degeneracy toward quantitative estimates of how observational uncertainties propagate into interior inferences.
~\cite{Otegi2020b} investigated several factors influencing the internal characterization of super-Earths and sub-Neptunes, including observational uncertainties, their position on the M-R diagram, the impact of additional constraints such as bulk composition and irradiation, and the role of model assumptions. They examined how the determination of internal structure depends on observational uncertainties across exoplanets with varying masses and radii. 

\begin{figure}
\centering
\includegraphics[width=0.95\linewidth]{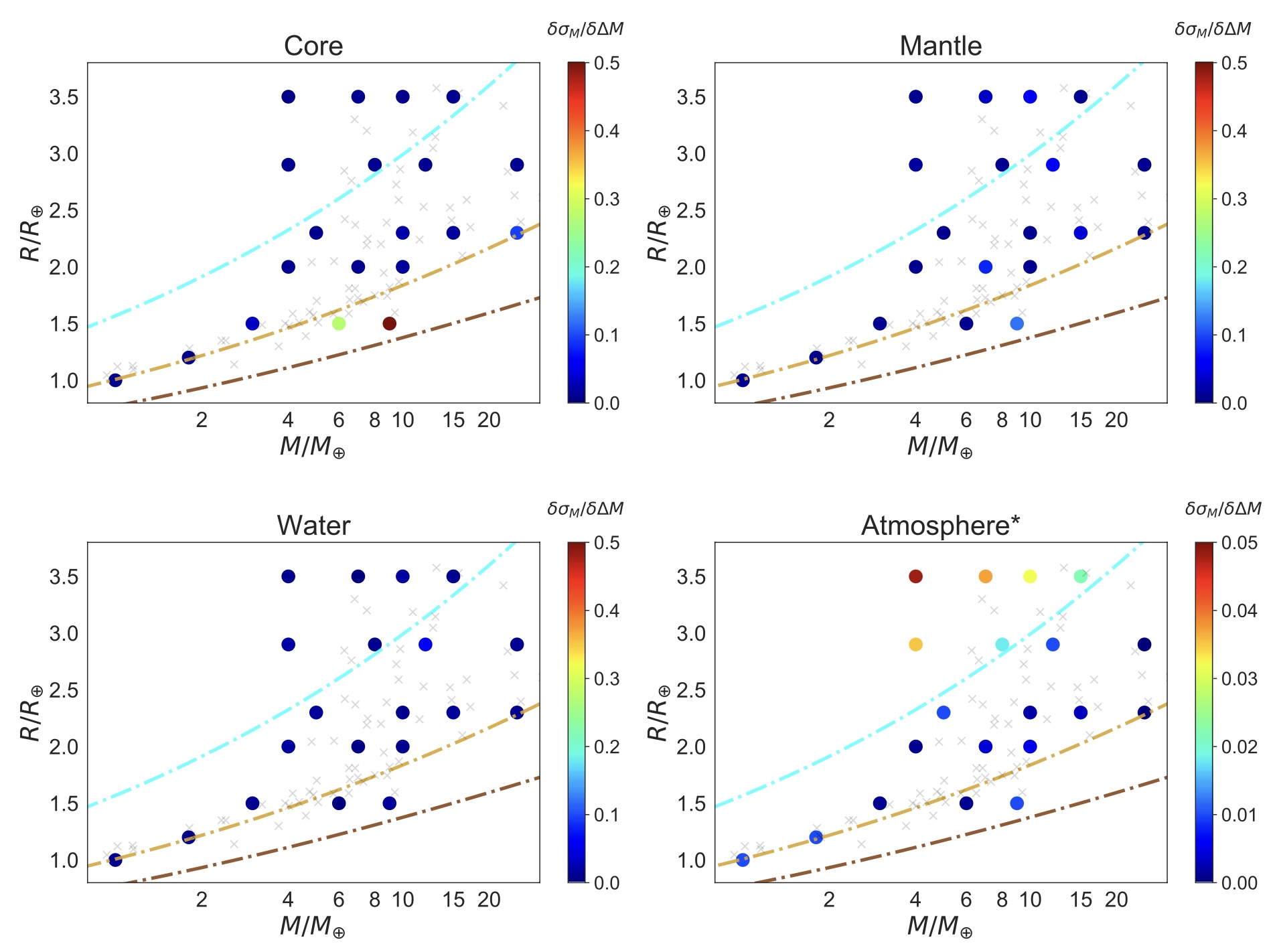}
\caption{\label{fig:otegi} Model parameters and synthetic planets $\Delta \sigma / \Delta M$. A higher value of $\Delta \sigma/ \Delta M$ indicates that a decrease of the observational uncertainty provides additional information of the internal parameter. The crosses represent the observed planets. Composition lines of pure-iron (brown), Earth-like planet (light brown), and water ice (blue) are also plotted (Figure from \citet{Otegi2020b}).
}
\end{figure}

It was found that reducing observational uncertainties for planets below the Earth-like composition line leads to a more precise determination of core mass. In this density regime, planets lack a significant gaseous envelope and require a high iron content. Consequently, variations in mass or radius distributions directly impact core mass estimates, as no other interior layer can account for such high bulk densities. 
Similarly, for planets above the pure-water composition line, improved observational precision enhances the determination of atmospheric mass. The underlying mechanism is analogous to that of dense planets: for a given mass, low-density planets require thick gaseous envelopes to match the observed radius, thereby reducing degeneracy in interior modeling. However, the transition between rocky and volatile-rich planets remains highly degenerate. They also suggested that using stellar Fe/Si and Mg/Si abundances as proxies for bulk planetary compositions does not consistently provide additional constraints on the internal structure.
Next, they explored how uncertainties in the planetary temperature gradient contribute to radius uncertainties. They showed that small variations in temperature or entropy profiles result in radius variations that are comparable to the observational uncertainties. This suggests that as observational precision improves, uncertainties associated with model assumptions may eventually become more influential in determining internal structure than observational uncertainties themselves. \\

\citet{2024MNRAS.530.3488P} suggested that precision in planet mass and radius remains the primary driver for the characterization of intermediate/small-mass planets: improved (few \%) measurements shrink allowed ranges for core size and volatile fraction, but there are diminishing returns depending on where a planet sits in mass–radius space (dense, rocky vs. volatile-rich regimes). At the same time, using host-star refractory abundances (Fe, Si, Mg) as priors for bulk composition substantially can reduce the  degeneracy in mantle/core partitioning and compositional solutions compared with the M-R relation alone. Therefore, combining precise mass and radius measurements together with stellar abundances yields much tighter interior constraints than any one measurement by itself \citep{dorn2015can,Luque2022}.

In addition, machine-learning surrogates and mixture-density networks such as  ExoMDN \citep{baumeister2020machine}  speed up probabilistic interior retrievals and make it easier to explore observational-error-driven posterior spreads. Updated mass–radius population studies and re-analyses revise the boundaries between rocky, water-rich, and gas-rich regimes, which in turn affects how a given M–R measurement is interpreted for interior composition. 
\citep{2024A&A...692A.113W} introduced the HADES model that couples an atmosphere and interior self-consistently, using radiative-convective atmospheric calculations plus ab initio interior equations of state. This work applies it to exoplanets like WASP-39 b and 51 Eri b; for WASP-39 b, spectroscopic constraints allow the atmospheric metallicity and intrinsic temperature  to be much better constrained than from mass and radius alone. It was shown that when using spectrum data, the derived metallicity and core mass shift outside the 68\% posterior region one would get using only mass and radius, so atmosphere model and spectrum data can reduce degeneracies significantly (see also \citet{2024A&A...688A..60A}).

Finally, we note that non-planetary measurement uncertainties can be as important as formal mass/radius errors so comprehensive inference now routinely folds in stellar and observational systematics when reporting interior constraints \citep{muller2020theoretical}. \\

{\it Bayesian inference analysis of planetary interiors} \\

A thorough characterization of exoplanet interiors requires to quantify confidence regions of interior parameters as parameter degeneracy is generally large. Within $PlanetS$, a generalized Bayesian inference analysis \citep{dorn2015can, dorn2017bayesian,Haldemann2024} has been developed and then employed in many observational studies (40+) to estimate interior structure and compositions (e.g., core-mantle ratios, water budgets, rock composition, gas mass fraction) of newly discovered exoplanets.  It has also inspired other groups to create similar tools for interior characterization \citep[e.g.,][]{huang2022magrathea, baumeister2020machine,brugger2017constraints}. 

To reduce the generally large degeneracy, constraints in addition to mass and radius can be used, that were previously not considered in characterization studies. Some examples include: Stellar abundance proxies \citep{dorn2015can} help to better constrain the rocky interior, which was adapted by many groups \citep[e.g.,][]{unterborn2017inward,brugger2017constraints,schulze2021probability,plotnykov2020chemical}. How stellar abundances are correlated to planetary rock compositions and how much it deviates from a direct 1:1 ratio due to planet formation processes is an ongoing debate  \citep{Adibekyan2021, schulze2021probability}.
Further, age and stellar insolation allow to identify and characterize atmospheres of secondary origin (outgassed), when the loss of a primordial hydrogen is considered \citep{dorn2018secondary} or when observations on the exosphere exist \citep{Bourrier2018b}. Deep volatile reservoirs likely decrease the efficiency of  hydrogen loss \citep{Dorn&Lichtenberg, luo_interior_2024}, however further investigations are needed to quantify this.
 
Given constraints from tidal heating, it is possible to identified potential magma ocean worlds HD~219134~b \citep{ligi2019stellar} orbiting interior of a colder twin super-Earth, an ideal target for comparative exoplanetology. Also, using the condensation sequence of proto-planetary gas disks \citep{dorn2018new}, a potential new class of exoplanets have been identified that formed from high-temperature condensates. 
This highlights that constraints in addition to mass and radius are key to unravel the diversity of planet interiors, as M-R data alone only provide limited information \citep{Otegi2020b}. 

Besides individual planets, multi-planetary systems allow to make data types available that are specific to multi-planetary systems only  \citep{dorn2018interior,dorn2018new}. For example, the inferred water mass fractions for any TRAPPIST-1 planet also depends on the M-R data of its neighboring planets \citep{dorn2018interior}, as M-R data are correlated across neighboring planets. Further, multi-planetary systems offer to place constraints on bulk composition across neighboring planets \citep{Agol2021}, which can improve estimates on water budgets \citep{dorn2018new} or core-mass fractions \citep{dorn2018interior}. 

Recent developments on modeling deep water reservoirs for warm/hot exoplanets \citep{Dorn&Lichtenberg} are fundamental steps for the development of new generation models. Models accounting for chemical and compositional coupling between the atmosphere and the deep interior can correctly infer bulk water budgets which can be an order of magnitude higher compared to commonly used models, where distinct and chemically inactive layers are assumed. With the thousands of new exoplanet targets to come from PLATO \citep{Rauer2025}, accurate models are the backbone of their meaningful interior characterization.

Planets evolve in time, they cool, ingas, and outgas on long time-scales. Planets in stagnant-lid regime regardless of their diversity in rock compositions show efficient outgassing only for small planet masses below 6-7 \ME \citep{dorn2018outgassing}, however accurately treating melting and crust production indicates outgassing for larger planet masses (Tackley et al. in prep). We have also demonstrated that the composition of the atmosphere depends on the melt fraction of the mantle. Hence very young terrestrial planets or those that are efficiently heated at depth have carbon-rich (reduced) atmospheres while water-dominated (oxidised) atmospheres evolve when planets have cooled enough that melt fractions are small \citep{bower2019linking,bower2021retention}. The atmosphere composition is also influenced by the redox state of a planet; reduced mantles preferentially outgas H$_2$ and CO, while oxidised mantles outgas H$_2$O and CO$_2$ \citep{ortenzi2020mantle}. Hence, atmosphere composition may inform us about the composition of the deep interior and its evolution. \\

{\it Linking planetary and stellar compositions} \\

Planets and the host star in one system are born from the same molecular cloud, resulting in a shared chemical origin of them. Gas-dust fractionation in a protoplanetary disk and subsequent dynamical processes may cause the bulk composition of (particularly rocky) planets to deviate from that of their host star. This is particularly true for highly and moderately volatile elements -- such as, carbon, oxygen, sulfur, and sodium, which are subjective to thermal processing, as observed in both the solar system rocky bodies (\cite{Palme2014, Wang2019, Braukmuller2019, Sossi2022}) and exoplanet systems [particularly with polluted white dwarfs; \cite{Harrison2021, Xu2021, Aguilera-Gmez2025}]. However, by analogy with the compositional differences between Earth, Mars and the Sun, the relative proportions of refractory (rock-forming) elements between rocky exoplanets and their host stars are expected to be within $\sim$10-20 \% \cite{Wang2018, Wang2019, Khan2022}. Based on a sample of high-precision host stellar abundances of planet-hosting stars and a stoichiometric disk compositional model, \cite{Adibekyan2021} suggested a potential deviation for refractory elements as well between super-Earths and their host stars.\\ 

Nevertheless, linking planets to stars points to an imperfect, yet extremely valuable way of constraining the diversity of exoplanets across rocky, super-Earth, and sub-Neptune regimes. The foremost advantage of adopting stellar abundances as proxies for the bulk compositions of those (potential) planets around them is that they are measurable with stellar photospheric spectroscopy, thus providing a broad observationally-informed bound on planetary composition \cite{Hinkel2018, Wang2022mnras, Spaargaren2020, Spaargaren2023}. \\


\section{Formation and evolution of super-Earths and sub-Neptunes}
\label{sec:evo}

Super-Earths and sub-Neptunes are the most common type of exoplanets discovered to date at orbital periods below 100 days. How did they form? What is the physical explanation for the radius valley? Why do we not have any super-Earth or sub-Neptune in our Solar System? These fundamental questions have driven intense research within the international community and by $PlanetS$ researchers since the finding of the radius valley by \citet{fulton_california-kepler_2017}. \\


{\it The possible physical explanations of the radius valley} \\

The first models capable of reproducing the radius valley as a depletion at $1.5-2.0$ Earth radii in the theoretical size distribution of planets were pure evolution models. In these models, the protoplanetary disc is already dissipated and the planets are assumed as fully formed at the beginning of the simulations. The physics that drives the evolution are the cooling of the planets, and the erosion of the atmospheres due to some heat source. In the case of photoevaporation \citep[][]{Owen2017, jin2018compositional, Mordasini2020} the source of heat eroding the atmosphere is the irradiation from the central star, while in the case of core-powered mass-loss is the accretion heat slowly released from the core \citep{Ginzburg2018, gupta2019signatures}. Both mechanisms find that the resulting small planets are either fully stripped rocky cores (the super-Earths) or rocky cores surrounded by a H/He envelope which amounts to 1\% of the planet's mass and doubles the core size (the sub-Neptunes). Both models can reproduce the radius valley with the following two simple assumptions. i) Super-Earths and sub-Neptunes were the same class of planets after formation: a core of a given (single) material surrounded by a H/He envelope; ii) the core mass distribution was uni-modal, peaked at approximately 3~\ME \citep[Rayleigh distribution, see][]{Owen2017}.  
In the light of those assumptions, both scenarios showed that the correct location of the valley was only reproduced when the core composition was rocky. Icy cores or iron cores would yield a too large or too small radius valley, respectively. Such finding led to the conclusion that the majority of super-Earths and sub-Neptunes formed inside the water ice line \citep{Owen2017}. 

Nevertheless, formation models were telling a different story at that time. From a formation perspective, sub-Neptunes accrete their cores predominantly beyond the iceline, where there are more solids that allow the buildup of large cores. In addition, for Sun-mass stars, planets start to migrate efficiently once they reach approximately 1 \ME. Thus, formation simulations show that planets in the mass, size and orbital locations of today's sub-Neptunes, formed mostly in the icy regions of the disc and are thus, water-rich. This result is very robust regarding model ingredients, with both planetesimal- and pebble-based models predicting that sub-Neptunes should be typically water-rich \citep{Alibert13, Venturini2017, Bitsch18, Brugger20} due to the efficient type-I migration. Furthermore, when models attempted to form sub-Neptunes inside the iceline only, they could not reproduce the second peak of the bimodal size distribution because rocky cores do not grow enough to be able to hold a H/He atmosphere after $\sim$100 Myr of evolution with photoevaporation \citep{Venturini2020b}. 

\begin{figure}
    \centering
    \includegraphics[width=\linewidth]{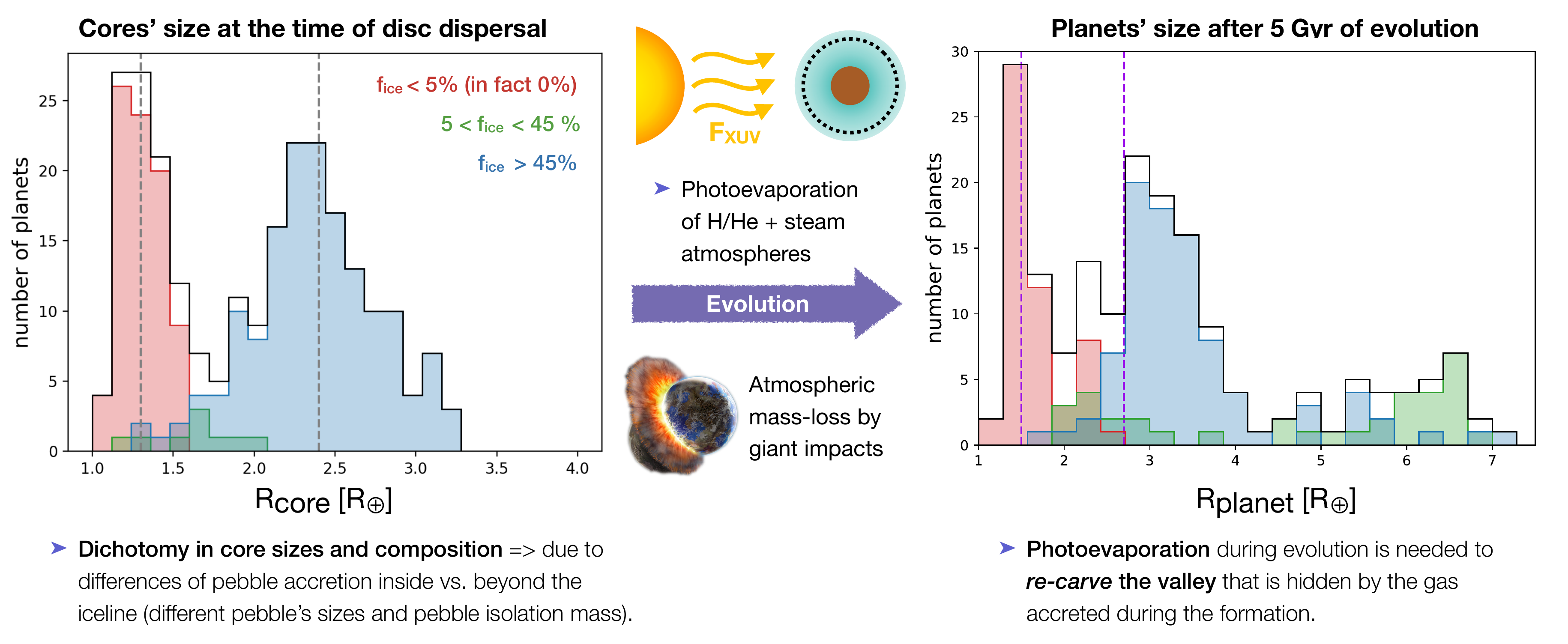}
    \caption{Sketch illustrating the explanation of the radius valley from \citet{Venturini2020a} (adapted from their paper). The left panel shows the distribution of the core sizes at the end of the formation phase (when the disc dissapears). The right panel shows the distribution of the size of the planets (cores+atmospheres) after 5 Gyr of evolution, after including atmospheric-mass erosion by photoevaporation and giant impacts.}
      \label{fig:rad_valley_venturini}
\end{figure}

\citet{Venturini2020a} showed that it is possible to reconcile formation and evolution models. They performed global formation and evolution simulations including the evolution of the gas and dust in the disc, pebble and gas accretion, planet migration, and planet cooling with photoevaporation during 5 Gyr after disc dispersal. They found that the differences in pebble size between rocky and icy pebbles naturally leads to two distinct cores' populations: icy and large versus rocky and small. This bimodality from birth naturally matches the location of the radius valley (see left panel of Fig.\ref{fig:rad_valley_venturini}). In reality, the cores bind some atmosphere during the accretion process, which blurs the primordial radius valley. Nevertheless, \citet{Venturini2020a} showed that after the disc dissipates, photoevaporation efficiently removes the gas from the planets whose cores straddle the radius valley, re-carving the valley in the correct location (see right panel of Fig.\ref{fig:rad_valley_venturini}). 
In this way, the radius valley naturally separates rocky super-Earths from water-rich sub-Neptunes. A key difference with the assumptions of pure evolution models from \citet{Owen2017} and \citet{Gupta2019}, is that the predicted core mass function which naturally stems from the formation simulations of \citet{Venturini2020a} is bimodal, both in core mass and in core composition (small rocky cores vs. larger water-rich cores with a water-to-rock ratio of typically 1).
In this way, \citet{Venturini2020a} showed that a consistent picture of formation and evolution can explain the populations of super-Earths and sub-Neptunes, predicting that sub-Neptunes should be predominantly water-worlds. In the view of this global formation and evolution model, the conclusion from pure evolution models was incorrect due to assuming a uni-modal core mass distribution, and due to not considering water in the form of steam (and prone to be evaporated) atop the rocky cores. Later, \citet{Burn24} also concluded that sub-Neptunes should be water-worlds by a population synthesis approach based on planetesimal accretion. The main effect separating rocky vs. water-worlds in that case is not a bi-modal core mass function as in \citet{Venturini2020a}, but the effect of the steam atmospheres, which puffs up the radii of the water-worlds. In that regard, it is important to highlight a fundamental tool developed during PlanetS: the compilation of a table of equations of state for water (AQUA) by \citet{haldemann2020aqua}. Thanks to that new tool, the steam atmospheres of sub-Neptunes could be properly modelled in \citet{Burn24} and \citet{Venturini2024}. \\
\vspace{-0.7cm}

{\it The dependence of the radius valley on stellar and orbital properties} \\

The radius valley has also a dependence on stellar mass and orbital period (see Sect. 12.3). This provides additional demographics constraints to benchmark planet formation and evolution models. \citet{Venturini2024} expanded the 2020 model to stellar masses ranging from 0.1 to 1.5 solar masses. They found that the radius valley fades when moving towards M dwarfs. This is the result of the dependence of the threshold for type I migration with stellar mass \citep{Burn21}. The lower the stellar mass, the lower the planet mass threshold to undergo type I migration \citep{Paardekooper2011}. Thus, around M dwarfs, sub-Earth-mass icy planets can reach close-in orbits, filling the radius valley \citep{Venturini2024}. The models of \citet{Venturini2024} also find a slope of the radius valley with stellar mass in agreement with observations. For the radius-period slope, they find a negative slope in the case of FGK stars, and a flat slope for M-dwarfs, trends that also follow the observed ones. 

Further theoretical work is needed to better reproduce the orbital period distribution of super-Earths and sub-Neptunes. In current global formation and evolution models type-I migration is so efficient that planets remain closer-in than observed \citep{Burn24}. This might hint towards the importance of disc winds in the evolution of protoplanetary disc. In those models, the surface density of gas as a function of orbital distance is less steep, which would naturally reduce type-I migration rates. 

Besides M-dwarfs, another type of planet-hosting stars in which is important to improve the characterisation and formation models of super-Earths and sub-Neptunes is binary stars. Half of the solar-type stars in our galaxy have a stellar companion \citep{Raghavan2010}, and the number of detected planets orbiting  binaries is increasing steadily, particularly S-type planets (which orbit one of the two stars in a binary). 
\citet{Sullivan24} performed the first demographics analysis of super-Earths and sub-Neptunes in S-type binary systems using Kepler data cross-matched with direct imaging follow-up (to identify the binaries). They found that the peak of sub-Neptunes is suppressed for projected binary separations below 100 au. They additionally found hints of the radius valley changing its location to lower radius for intermediate projected binary separations of $\sim$300-2000 au. Global formation models on the formation of S-type planets are being developed (Venturini et al. in prep., Nigioni et al. in prep), a necessary tool to confront theory with the inferred demographic trends. At the same time, further efforts should be made to improve the detection and characterization of planets orbiting binary stars.

The coherent picture that has emerged until now for the origin of super-Earths and sub-Neptunes suggests that sub-Neptunes can be considered as scaled-down versions of the ice giants, in the sense that both planet types should have formed beyond the iceline and be therefore rich in volatiles. 
Nevertheless, ice giants and sub-Neptunes have very different equilibrium temperatures and are thus expected to have very different interior structures (see Section on interiors).

Solving the riddle regarding the true composition of sub-Neptunes is a key future endeavor, which will allow us to  distinguish between different formation/evolution scenarios. One way to achieve this is by measuring atmospheric abundances with JWST or Ariel, and combine them with interior structure modeling to set constraints in the bulk amount of water of sub-Neptunes. Another possibility is to measure planetary radii at different ages with PLATO, because the differences in the cooling rates between water-rich and H/He planets makes the planet sizes evolve differently \citep{alibert2016constraining}.\\

{\it Why are not there any super-Earths or sub-Neptunes in the Solar System?} \\

Regarding the question of the lack of super-Earths and sub-Neptunes in the Solar System, it is broadly accepted that this is related to the early formation of Jupiter \citep{Izidoro15, Lambrechts19}. In the context of pebble-based models, the early formation of Jupiter would have suppressed the flux of pebbles drifting from the outer disc, depleting the inner solar disc from solids to form large enough cores \citep{Lambrechts19}. In this regard, a negative correlation between the presence of super-Earths/sub-Neptunes and outer giants should be expected. In the context of planetesimal-based models, population synthesis results showed that while the general occurrence of super-Earths/sub-Neptunes declines with the formation of a cold giant, a positive correlation should exist between the presence of dry super-Earths and cold giants \citep{Schlecker2021}. This is because the formation of an outer giant does not preclude an inner rocky embryo from growing close to the star and accreting dry planetesimals, but it prevents icy planets from migrating to the inner regions of the disc \citep{Schlecker2021}. Observationally, it is still debated wether there is a correlation or not between the presence of inner super-Earths/sub-Neptunes and outer giants \citep{Zhu18, Rosenthal22, Bonomo23}. Recent work points towards a positive correlation only in the case of high-metallicity stars \citep{Bonomo25}. Deciphering these type of correlations is important to understand our own Solar System in the context of Exoplanets.



\section{Open questions and future prospects}


Our understanding of the formation and characterization of super-Earths and sub-Neptunes has improved significantly in the last decade, with key contributions from $PlanetS$. Yet, their diversity in mass, radius, and inferred bulk composition points to a complex formation history, atmospheric loss, and late-stage evolution processes. Comparative planetology studies, bolstered by improved M-R relations, have begun to disentangle the relative roles of core mass, envelope accretion, and evolutionary pathways.
Despite this progress, fundamental questions remain. The transition between rocky super-Earths and volatile-rich sub-Neptunes, the prevalence of water worlds, and the atmospheric compositions of these planets are still poorly constrained. \\

The efforts to characterize planets is ongoing and thanks to current (and future) space missions and ground-based observations, exoplanets can now be characterized in greater detail. However, just like Uranus and Neptune, intermediate-mass planets are not well-understood. Often these planets are modeled by the scaling up of terrestrial planets or the scaling down of giant planets, but our experience from the Solar System planets have taught us that these planets correspond to a unique planetary class that deserves an independent modeling approach \citep[e.g.,][]{HelledFortney2020}. Currently, intermediate-mass exoplanets are modeled in a traditional manner where the interior is assumed to consist of distinct layers (a rocky core, layer composed of ices, and H-He atmosphere with various enrichments) and the temperature profile is assumed to be adiabatic \citep[e.g.,][and references therein]{Otegi2020b}. However, it is unclear whether the interiors of these planets consist of distinct layers, how the heat is transported, and what the dominating elements in their interiors are. Several efforts to model the interiors of such planets have been made.

JWST is beginning to demonstrate its ability to probe the atmospheric composition of super-Earths and sub-Neptunes, confirming their broad compositional diversity \citep[e.g.,][]{Benneke2024,Piaulet2024,Alderson2024,Gressier2024,Schmidt2025,Ohno2025,Malsky2025,Ahrer2025}. Expanding such observations will be essential to further advance our understanding of these worlds. Future space-based missions such as ARIEL \citep{Tinetti2018}, along with ground-based facilities like ANDES@ELT \citep{Marconi2022}, will also dramatically improve spectroscopic characterization of their atmospheres, allowing measurements of molecular abundances, temperature structures, and cloud properties. In parallel, theoretical models must be further developed to interpret this rich data. The synergy between future observations and improved modeling efforts promises to transform our understanding of super-Earths and sub-Neptunes in the coming decade. \\

The primary goal of the PLATO (PLAnetary Transits and Oscillations of stars) mission \citep{Rauer2025}, which is set to be launched in end 2026, is to open a new avenue in exoplanetary science by detecting small-size exoplanets up to the temperate zone and characterizing their bulk properties. PLATO will provide the key information (radii, bulk densities, insolation, architecture as well as the age) needed to determine the habitability of these unexpectedly diverse new worlds. The first PLATO field (LOPS2) \citep{LOPS22025} was recently selected in the southern hemisphere with a significant overlap with the TESS South Continuous Viewing Zone \citep{Eschen2024}. The simulated PLATO planet yield \citep{Matuszewski2023} predicts that PLATO will detect several hundreds of temperate Earths, super-Earths, and sub-Neptunes, with one-third of them transiting stars brighter than V=11. The PLATO Ground-based Observation Program (GOP) oversees providing the complementary data fundamental to establish the nature of the transit events detected in the PLATO light curves, determining the best stellar parameters of the planet hosts, and providing precise mass for the validated candidates. \\

However, studies of exoplanetary systems suffer from a largely biased and incomplete view by probing mainly the inner part of the planetary system due to transit probability and due to the sensitivity of RV techniques to small orbital periods. This bias prevents us from having a global understanding of the architecture of the planetary systems which may reveal their history and evolution. The ESA space mission Gaia, which has been monitoring the entire Milky Way since 2014, is probing by astrometry the outer region of planetary systems. Gaia will be able to reveal cold giant planets (with mass larger than Saturn) that are beyond the ice line of all known planetary systems and/or transiting candidates \citep[e.g.,][]{Sozzetti2014,Perryman2014,Holl2023}. \\

The Nancy Grace Roman Space Telescope \citep{Spergel2015}, set to launch no later than May 2027, will complement PLATO and Gaia by probing the outer regions of planetary systems through microlensing and direct imaging. Roman will be sensitive to cold Super-Earths and sub-Neptunes beyond the snow line, filling the gap left by transit and RV surveys, and enabling the study of their occurrence, architecture, and atmospheric properties. By combining Roman’s census of wide-orbit planets with inner-system data, we will gain a more complete picture of planetary system formation and evolution. \\

Finally,  the influence of stellar multiplicity on planet formation remains a major open question. Approximately half of Sun-like stars belong to binary or multiple systems \citep{Raghavan2010}, and binarity is likely to affect both the formation and long-term evolution of planetary systems. Characterizing exoplanets in binaries is key to determining how a stellar or a brown dwarf companion influences protoplanetary disk and planetary formation and architecture. S-type planets - planets orbiting only one component of the binary system – are expected to form in systems with separations sufficient to sustain individual protoplanetary disks (PPDs) around each star \citep{Bate2000}. Iconic examples of S-type planets are Gamma-Cephei \citep{Thebault2004}, K2-136 \citep{K2-136}, and Kepler-444 \citep{Kepler444}. On the other hand, well known circumbinary planets (planets orbiting the two stars in a binary, also known as "P-type") are Kepler-16 \citep{Kepler-16} and and TOI-1338/BEBOP-1 \citep{Standing23}. The study of exoplanets in relatively close visual binaries so far remains limited due to observational challenges, such as dilution effects in photometry, leading to an incomplete understanding of planet occurrence in binaries. \\

In summary, while substantial progress has been made in understanding super-Earths and sub-Neptunes, the next generation of space missions and ground-based facilities, alongside theoretical advancements, will be instrumental in resolving the remaining mysteries surrounding these ubiquitous and diverse planets.

\bibliographystyle{spbasic}
\footnotesize
\bibliography{bib.bib}

\end{document}